\documentclass[aps,prd,groupedaddress,showpacs,showkeys]{revtex4}
\usepackage{latexsym,epsfig,amsmath,amssymb}
\usepackage{amssymb}
\usepackage{amsmath}
\usepackage{amsfonts}
\usepackage{epsfig} 
\usepackage{verbatim} 
\newcommand{\seq}{\begin{subequations}}
\newcommand{\sen}{\end{subequations}}
\newcommand{\eq}{\begin{eqnarray}}
\newcommand{\en}{\end{eqnarray}}
\newcommand{\ra}{\rangle}

\def\lc{\Lambda_c(2940)^+}
\def\lcp{\Lambda_c(2286)^+}

\def\L2{\Lambda^2}

\begin{document}
\hfill KEK-TH-1444 
\title{Strong three-body decays of $\Lambda_c(2940)^+$ 
in a hadronic molecule picture\\}  
\author{Yubing Dong$^{1,2,3}$, 
        Amand  Faessler$^3$,   
        Thomas Gutsche$^3$, 
        S. Kumano$^4$, 
        Valery E. Lyubovitskij$^3$\footnote{On leave of absence
        from Department of Physics, Tomsk State University,
        634050 Tomsk, Russia}
\vspace*{1.2\baselineskip}}
\affiliation{
$^1$ Institute of High Energy Physics, Beijing 100049, P. R. China\\ 
\vspace*{.4\baselineskip} \\
$^2$ Theoretical Physics Center for Science Facilities (TPCSF), CAS, 
Beijing 100049, P. R. China\\ 
\vspace*{.4\baselineskip} \\ 
$^3$ Institut f\"ur Theoretische Physik,  Universit\"at T\"ubingen,\\
Kepler Center for Astro and Particle Physics, \\ 
Auf der Morgenstelle 14, D--72076 T\"ubingen, Germany\\
\vspace*{.4\baselineskip} \\ 
$^4$ 
KEK Theory Center, \\
Institute of Particle and Nuclear Studies, \\ 
High Energy Accelerator Research Organization (KEK), \\
and Department of Particle and Nuclear Studies, \\
Graduate University for Advanced Studies, \\
1-1, Ooho, Tsukuba, Ibaraki, 305-0801, Japan\\}

\date{\today}

\begin{abstract} 
The $\Lambda_c(2940)^+$ baryon with quantum numbers
$J^P = \frac{1}{2}^+$ is considered as a hadronic molecule
composed of a nucleon and $D^\ast$ meson.
We give predictions for the width of
the strong three-body decay processes  
$\lc\to \lcp\pi^{+}\pi^{-}$ and $ \lcp\pi^0\pi^0$ 
in this interpretation.
Upcoming experimental facilities like a Super $B$ factory at KEK 
or LHCb might be able to provide data on these decay modes. 

\end{abstract}



\maketitle

\section{Introduction}

The charmed baryon $\lc$ was originally observed 
by {\it BABAR}~\cite{Aubert:2006sp} 
and later on confirmed by the Belle Collaboration~\cite{Abe:2006rz} 
as a resonant structure in the final state 
$\Sigma_c(2455) \pi \to \Lambda_c \pi \pi$. 
Both collaborations deduce values for mass and width with
$m_{\Lambda_c} = 2939.8 \pm 1.3 \pm 1.0$ MeV,
$\Gamma_{\Lambda_c} = 17.5 \pm 5.2 \pm 5.9$~MeV 
({\it BABAR}~\cite{Aubert:2006sp}) and 
$m_{\Lambda_c} = 2938.0 \pm 1.3^{+2.0}_{-4.0}$ MeV,
$\Gamma_{\Lambda_c} = 13^{+8 \ + 27}_{-5 \ -7}$ MeV 
(Belle~\cite{Abe:2006rz}) which are consistent with each other.

Theoretical interpretations of this new charmed baryon resonance 
were already discussed in the literature (see e.g. the short overview 
in Ref.~\cite{Dong:2010xv}) including a conventional understanding in
different types of three-quark and quark-diquark 
models~\cite{Capstick:1986bm}-\cite{Valcarce:2008dr}. 
In Ref.~\cite{He:2006is} it was proposed that the $\lc$ is a hadron molecule,
where this state is regarded as a $D^{\ast 0} p$ configuration with 
spin--parity being 
$J^P = \frac{1}{2}^-$ or $\frac{3}{2}^-$. This interpretation is due 
to the fact that the $\lc$ mass is just a few MeV below the $D^{\ast 0} p$ 
threshold value and therefore strong coupling to this hadron channel 
is expected. It was also shown that the boson-exchange mechanism, 
involving the $\pi$, $\omega$ and $\rho$ mesons, can provide binding 
for such $D^{\ast 0} p$ configurations. 
But in a first variant of a unitary meson-baryon coupled channel 
model~\cite{GarciaRecio:2008dp} the $\lc$ cannot be identified with a 
dynamically generated resonance. Hence a possible binding of 
$D^{\ast 0} p$ remains to be examined. 

We also studied the structure of the $\lc$ as a possible molecular state  
composed of a nucleon and a $D^\ast$ meson within a formalism related to the 
compositeness condition~\cite{Dong:2009tg,Dong:2010xv}. 
We analyzed its two-body strong and radiative partial decay widths for 
the channels of $pD$, $\Sigma_c(2455)\pi$ and $\Lambda_c(2286)\gamma$. 
In case of the two-body strong decays we tested two different spin-parity 
assignments for the $\lc$: $J^P = \frac{1}{2}^+$ and $\frac{1}{2}^-$. 
It was found that for 
$J^P = \frac{1}{2}^+$ the sum of the three partial widths is consistent with 
present observation, while for $\frac{1}{2}^-$ a severe overestimate for the 
total decay width is obtained. Hence we concluded in~\cite{Dong:2009tg} 
that the choice of spin-parity 
$J^P = \frac{1}{2}^+$ is preferred in the molecular interpretation. 
Furthermore, the radiative decay $\Lambda_c(2940)^+ \to 
\Lambda_c(2286)^+ \gamma$ has also been estimated using the same 
approach~\cite{Dong:2010xv} assigning the $J^P = \frac{1}{2}^+$ 
spin-parity to the $\lc$. 

In this brief report we extend our previous analysis 
to estimate the two-pion decay channels of the $\Lambda_c(2940)^+$ as
$\lc\to \lcp\pi^{+}\pi^{-}$ or $\lc\to \lcp\pi^0\pi^0$. 
Although these two-pion decay modes of the $\lc$ have also been discussed 
in Ref.~\cite {He:2006is} no quantitative results were presented yet. 
This is because an unknown coupling constant for the vertex  $ND^*\Sigma_c$
occured in the considerations of Ref.~\cite {He:2006is}. 
However, a quantitative prediction for the three-body decay
widths of $\lc\to \Lambda_c(2286)^+ + 2\pi$ could be done 
using information about 
two-body decays $\lc\to\Sigma_c(2455)+\pi$ done in Ref.~\cite{Dong:2009tg} 
and would be helpful for a measurement at the 
upcoming experimental facilities like Belle II at a Super $B$ factory 
at KEK or with LHCb. 

In this article the strong three-body decays of the $\lc$ baryon will be
analyzed using the technique based on the compositeness 
condition~\cite{Weinberg:1962hj,Efimov:1993ei}  
for describing and treating composite hadron systems
as developed in 
Refs.~\cite{Dong:2009tg},\cite{Faessler:2007gv}-\cite{Dong:2008mt}. 
In particular, 
in~\cite{Dong:2009tg,Faessler:2007gv,Dong:2010gu} 
recently observed unusual hadron states (like $D_{s0}^\ast(2317)$, 
$D_{s1}(2460)$, $X(3872)$, $Y(3940)$, $Y(4140)$, $Z(4430)$, 
$\Lambda_c(2940)$, $\Sigma_c(2800)$) were analyzed within the structure
assumption as hadronic molecules. 
The compositeness condition implies that the 
renormalization 
constant of the hadron wave function is set equal to zero or that the hadron 
exists as a bound state of its constituents. It was originally applied to 
the study of the deuteron as a bound state of proton and
neutron~\cite{Weinberg:1962hj} (see also Ref.~\cite{Dong:2008mt} for 
a further application of this approach to the case of the deuteron). 
Then it was extensively used
in low--energy hadron phenomenology as the master equation for the
treatment of mesons and baryons as bound states of light and heavy
constituent quarks (see e.g. Refs.~\cite{Efimov:1993ei,Anikin:1995cf}). 
By constructing a phenomenological Lagrangian including the 
couplings of the bound state to its constituents and the constituents 
to other final state particles we evaluated meson--loop 
diagrams which describe the different decay modes of the molecular states 
(see details in~\cite{Faessler:2007gv}). 

In the present paper we proceed as follows. In Sec. II we briefly review the 
basic ideas of our approach. Moreover, we consider the strong three-body 
decays of the $\lc$ baryon $\lc\to \Lambda_c(2286)^+ + 2\pi$ in this 
section. In the calculation of the three-body decay 
of the $\lc$ we consider two resonance contributions  
with the intermediate charmed baryon $\Sigma_c(2455)$ 
and $\rho^0$ meson. 
In Sec. III we present our numerical results, and, finally, 
in Sec. IV a short summary. 

\section{Approach}

Here we briefly discuss the formalism for the study of the composite 
(molecular) structure of the $\lc$ baryon. 
In the following calculation we adopt spin and parity quantum numbers
$J^P = \frac{1}{2}^{+}$ for the $\lc$, which is consistent with the 
observed strong decay width of the $\lc$ obtained in a hadronic molecule 
interpretation~\cite{Dong:2009tg}. Following the original suggestion of
Ref.~\cite{He:2006is} we consider 
this new baryon resonance as a superposition of molecular 
$p D^{\ast 0}$ and $n D^{\ast +}$ components with the adjustable mixing 
angle $\theta$: 
\eq\label{Xstate}
|\lc\ra =   \cos\theta \ | p D^{\ast 0} \ra \ 
        + \ \sin\theta \ | n D^{\ast +} \ra 
\, . 
\en 
The values $\sin\theta = 1/\sqrt{2}$, $\sin\theta = 0$ or  
$\sin\theta = 1$ correspond to the cases of ideal mixing, of 
a vanishing $n D^{\ast +}$ or $p D^{\ast 0}$ component, respectively. 
Since the observed mass value of the $\lc$ with
$m_{D^{\ast 0}} + m_p - m_{\lc} =  5.94$ MeV and
$m_{D^{\ast +}} + m_n - m_{\lc} = 10.54$ MeV lies closer to the
$p D^{\ast 0}$ than to the $n D^{\ast +}$ threshold, we might expect
that the $|p D^{\ast 0}\ra$ configuration is the leading component.
In this case the mixing angle $\theta$ should be relatively small 
and therefore we will vary its value from 0 to 25$^0$.

Our approach is based on an effective interaction Lagrangian describing 
the coupling of the $\lc$ to its constituents. We use a construction for 
the $\lc$ in analogy to mesons consisting of a heavy quark and a light 
anti-quark, i.e. the heavy $D^\ast$ meson sets the center of mass 
of the $\lc$ while the light nucleon moves around the $D^\ast$. 
The distribution of the nucleon relative to the $D^\ast$ meson 
is described by the correlation function $\Phi(y^2)$ depending on 
the Jacobi coordinate $y$. The simplest form of such a Lagrangian reads 
\eq\label{Lagr_Lc} 
\hspace*{-.2cm} {\cal L}_{\Lambda_c}(x) &=& 
\bar\Lambda_c^+(x) \, \gamma^\mu \int d^4y \, \Phi(y^2) \, 
\Big( g_{\Lambda_c}^0 \, 
\cos\theta \,D^{\ast 0}_\mu(x) \, p(x+y) 
    + g_{\Lambda_c}^+ \, 
\sin\theta \,D^{\ast +}_\mu(x) \, n(x+y) \Big) 
\ + \ {\rm H.c.} 
\,, 
\en 
where $g_{\Lambda_c}^+$ and $g_{\Lambda_c}^0$ are the coupling constants of 
$\lc$ to the molecular $nD^{\ast +}$ and $pD^{\ast 0}$ components. Here we 
explicitly include isospin breaking effects by taking into account the 
neutron-proton and the $D^{\ast +} - D^{\ast 0}$ mass differences. 
Note that in our previous analysis~\cite{Dong:2009tg} of strong two-body 
decays we restricted to the isospin 
symmetric limit. A basic requirement for the choice of an explicit form of 
the correlation function $\Phi(y^2)$ is that its Fourier transform vanishes 
sufficiently fast in the ultraviolet region of Euclidean space to render 
the Feynman diagrams ultraviolet finite. We adopt a Gaussian form for the 
correlation function. The Fourier transform of this vertex is given by 
\eq 
\tilde\Phi(p_E^2/\Lambda^2) \doteq \exp( - p_E^2/\Lambda^2)\,, 
\en 
where $p_{E}$ is the Euclidean Jacobi momentum. Here, 
$\Lambda$ is a size parameter characterizing 
the distribution of the nucleon in the $\lc$ baryon, which also leads to a 
regularization of the ultraviolet divergences in the 
Feynman diagrams. From the analysis of the strong two-body decays of the 
$\lc$ baryon we found that $\Lambda \sim 1$ GeV ~\cite{Dong:2009tg}. 
The coupling constants $g_{\Lambda_c}^+$ and $g_{\Lambda_c}^0$ are 
determined by the  compositeness condition~\cite{Weinberg:1962hj, 
Efimov:1993ei,Anikin:1995cf, Dong:2009tg,Faessler:2007gv}. It implies 
that the renormalization constant of the hadron wave function is set equal 
to zero with:
\eq\label{ZLc} 
Z_{\Lambda_c} = 1 - \Sigma_{\Lambda_c}^\prime(m_{\Lambda_c}) = 0 \,.
\en
Here, $\Sigma_{\Lambda_c}^\prime(m_{\Lambda_c})$   
is the derivative of the $\lc$ mass operator (see details 
in ~\cite{Dong:2009tg}).  

\begin{figure}
\centering
\includegraphics [width=6.5cm]{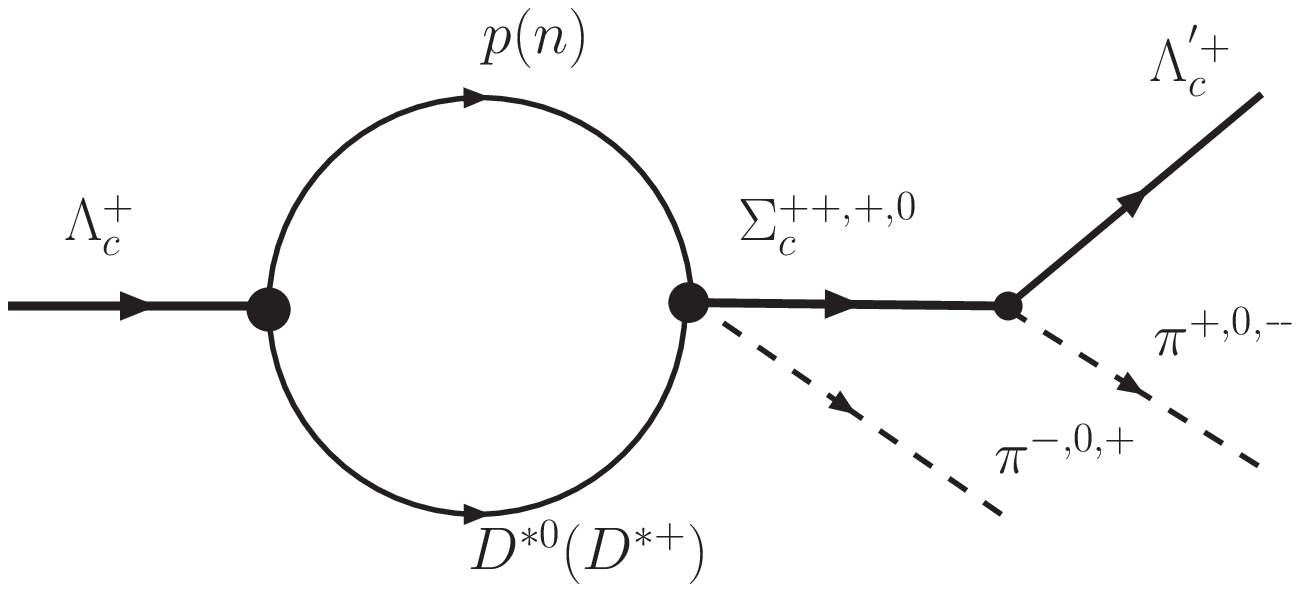}
\hspace{1cm} 
\includegraphics [width=6.5cm]{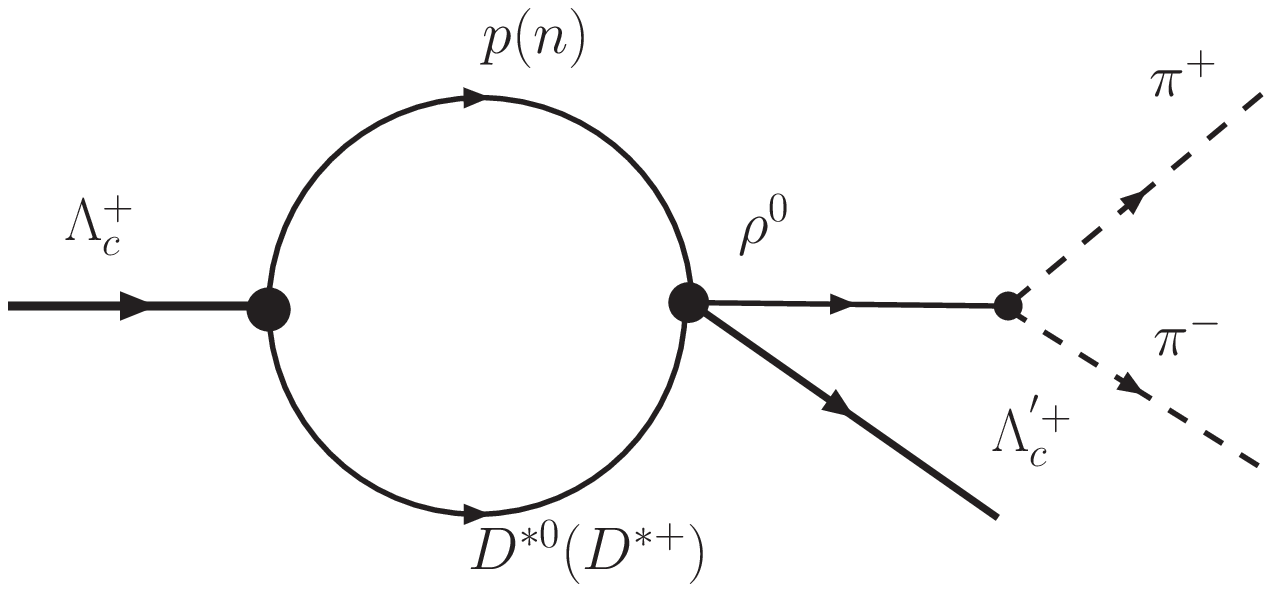}

\vspace*{.25cm}
\hspace*{-1.75cm} (a) \hspace*{7.25cm} (b) 
\caption{Diagrams contributing to the $\lc \to \lcp +2\pi$ decay}
\end{figure}

In the calculation of the three-body decay 
$\lc \to \lcp + 2\pi$ we consider two resonance contributions: 
with the intermediate charmed baryon $\Sigma_c(2455)$ 
[see Fig.1(a)] and for the $\rho^0$ meson [see Fig.1(b)] in the transition. 
Note, the diagram in Fig.1(b) only contributes to the process with 
a charged $\pi^+\pi^-$ pair in the final state. 
The full  matrix element of the three-body decay $\lc \to \lcp +2\pi$ 
is calculated using a phenomenological Lagrangian formulated in terms 
of hadronic degrees of freedom with: 
\eq\label{L_eff_tot}  
{\cal L}_{\rm eff} = {\cal L}_{\Lambda_c} + {\cal L}^{a} +  {\cal L}^{b} \, . 
\en 
The Lagrangian contains the following terms --- the coupling of $\lc$ with
the constituents 
$({\cal L}_{\Lambda_c})$, the terms ${\cal L}^{a}$ and ${\cal L}^{b}$ 
describing the two-step transitions of the $\lc$ constituents to 
the final state of Figs.1(a) and 1(b), respectively.  
In particular, the term 
\eq\label{L_eff_a} 
{\cal L}^{a} = {\cal L}_{\pi D^\ast N \Sigma_c} 
+ {\cal L}_{\pi\Sigma_c\Lambda_c'} 
\en 
contains the $\pi D^\ast N \Sigma_c$ and 
$\pi\Sigma_c\Lambda_c'$ couplings. These vertices are deduced from 
the SU(4) symmetric Lagrangians 
originally derived in~\cite{Okubo:1975sc} and then  
extensively employed in our formalism in 
Refs.~\cite{Dong:2009tg,Dong:2010gu,Dong:2010xv}:   
\eq\label{piDB1B2}
{\cal L}_{\pi^-D^{*0}p\Sigma^{++}_c}
&=&\Bigg [\frac14(g_1+g_2)-\frac32 g_3 \Bigg ]
\bar{\Sigma}^{++}_c\pi^-i\gamma^{\mu}\gamma_5pD^{*0}_{\mu}
\ + \ {\rm H.c.}\,, \nonumber \\
{\cal L}_{\pi^-D^{*+}n\Sigma^{++}_c}
&=&-\frac32 g_3
\bar{\Sigma}^{++}_c\pi^-i\gamma^{\mu}\gamma_5nD^{*+}_{\mu}
\ + \ {\rm H.c.}\,, \nonumber \\
{\cal L}_{\pi^0D^{*0}p\Sigma^{+}_c}
&=&\frac{1}{2}\Bigg [\frac14(g_1+g_2)-3g_3\Bigg ]
\bar{\Sigma}^+_c\pi^0i\gamma^{\mu}\gamma_5pD^{*0}_{\mu}
\ + \ {\rm H.c.}\,, \nonumber \\
{\cal L}_{\pi^0D^{*+}n\Sigma^{+}_c}
&=&\frac{1}{2}\Bigg [\frac14(g_1+g_2)-3g_3\Bigg ]
\bar{\Sigma}^+_c\pi^0i\gamma^{\mu}\gamma_{5}nD^{*+}_{\mu}
\ + \ {\rm H.c.}\,, \nonumber \\
{\cal L}_{\pi^+D^{*0}p\Sigma^{0}_c}
&=&-\frac32 g_3
\bar{\Sigma}^0_c\pi^+i\gamma^{\mu}\gamma_5pD^{*0}_{\mu}
\ + \ {\rm H.c.}\,, \nonumber \\
{\cal L}_{\pi^+D^{*+}n\Sigma^{0}_c}
&=&\Bigg [\frac14(g_1+g_2)-\frac32 g_3\Bigg ]
\bar{\Sigma}^0_c\pi^+i\gamma^{\mu}\gamma_5nD^{*+}_{\mu} 
\ + \ {\rm H.c.}\,, 
\en
and 
\eq\label{piB1B2}
{\cal L}_{\pi\Sigma_c\Lambda_c^{'+}}
=-\frac12\sqrt{\frac32}(g_2'-\frac12g_1')
\bar{\Lambda}_c^{'+}i\gamma^5\pi\Sigma_c+ \ {\rm H.c.}\, .
\en 
The effective couplings $g_i$ and $g_i'$ are fixed 
as~\cite{Dong:2009tg,Dong:2010gu,Dong:2010xv} 
\eq
& &g_1=0,~~~ g_2=-\frac{2}{5F_\pi}g_Ag_{\rho\pi\pi},~~~
g_3=-\frac{2}{3F_\pi}g_Ag_{\rho\pi\pi}\,, \nonumber\\
& &g'_1=0, ~~~~~g'_2=-\frac45\sqrt{2}g_{\pi NN} \,.
\en 
Here $F_\pi=92.4MeV$ is the pion decay constant, 
$g_{\pi NN}=13.2$ is the pion-nucleon coupling constant, 
$g_A=1.2695$ is the nucleon axial charge, 
$g_{\rho\pi\pi} = 6$ is the coupling of the $\rho$ meson 
to pions. We also introduce the notation 
$\Lambda_c^{'+}$ for the $\lcp$ baryon. 

The effective Lagrangian ${\cal L}^{b}$ involved in the calculation 
of the diagram Fig.1(b) also contains two terms: 
\eq\label{L_eff_b} 
{\cal L}^{b} = {\cal L}_{\rho D^\ast N \Lambda_c'} 
+ {\cal L}_{\rho\pi\pi} \,.  
\en 
Here, ${\cal L}_{\rho\pi\pi}$ is the effective Lagrangian of the 
$\rho\pi\pi$ coupling having the standard form 
\eq
{\cal L}_{\rho\pi\pi}=g_{\rho\pi\pi}\rho^{\mu}_{k}
\pi_i\partial_{\mu}\pi_j\epsilon_{ijk},
\en 
where $i,j,k$ represent the isospin indices. The Lagrangian 
${\cal L}_{\rho D^\ast N \Lambda_c'}$ can be derived using the procedure
suggested in Ref.~\cite{Oh:2007ej}. In particular, we start with 
the non-minimal (tensorial) $ND^*\Lambda_c'$ coupling 
\eq 
{\cal L}_{D^\ast N\Lambda_c'} = - g_{D^\ast N\Lambda_c'} \, 
\kappa_{D^\ast N\Lambda_c'} \, \bar N \, \sigma^{\mu\nu} \,  
\partial_\nu \, D^\ast_\mu \Lambda_c' \ +  \ {\rm H.c.}\,, 
\en 
where the couplings $g_{D^\ast N\Lambda_c'}$ and 
$\kappa_{D^\ast N\Lambda_c'}$ are fixed as~\cite{Dong:2010xv,Oh:2007ej}:  
\eq 
g_{D^\ast N \Lambda_c'}= - \frac{\sqrt{3}}{2}g_{\rho\pi\pi}\,, 
\quad 
\kappa_{D^\ast N\Lambda_c'}= 2.65 \,. 
\en 
In a next step we gauge the derivative acting on the $D^\ast$ meson 
by introducing the $\rho^0$-meson field as: 
\eq 
\partial_\nu D^\ast_\mu \to 
(\partial_\nu - \frac{i}{2} g_{\rho\pi\pi} \rho^0_\nu) \, D^\ast_\mu  \, .
\en 
It finally results in the  $\rho^0D^\ast N\Lambda_c'$ coupling: 
\eq
{\cal L}_{\rho D^\ast N\Lambda_c'}=
\frac{g_{\rho D^\ast N \Lambda_c'}}{2M_N}\bar{N}D^{*+}_{\mu} \, 
i\sigma^{\mu\nu} \, \rho_{\nu} \, \Lambda_c^{'+} \ + \ {\rm H.c.}\,,
\en
where $g_{\rho D^\ast N \Lambda_c'} = g_{\rho\pi\pi} g_{D^\ast N\Lambda_c'} 
\kappa_{D^\ast N\Lambda_c'}/2$. 
 
In the evaluation of the two diagrams of Fig.~1 we use the
standard free propagators for the intermediate particles:
\eq
iS_N(x-y)&=&\left<0|TN(x)\bar N(y)|0\right>=\int\frac{d^4k}{(2\pi)^4i}\,
e^{-ik(x-y)} S_N(k),\quad \nonumber \\
S_N(k)&=&\frac{1}{m_N - \not\! k-i\epsilon}
\en
for the nucleons and
\eq
iS_{D^\ast}^{\mu\nu}(x-y)&=&
\left<0|TD^{\ast\,\mu}(x)D^{\ast\,\nu\,\dagger} (y)|0\right>=
\int\frac{d^4k}{(2\pi)^4i}\,e^{-ik(x-y)} S^{\mu\nu}_{D^\ast}(k)\,,
\quad\nonumber \\
S^{\mu\nu}_{D^\ast}(k)&=&
\frac{-g^{\mu\nu}+k^\mu k^\nu/m_{D^\ast}^2}{m_{D^\ast}^2-k^2-i\epsilon}
\en
for the $D^\ast$ vector mesons. 
The contributions of the intermediate resonance states, the $\Sigma_{c}(2455)$ 
baryon and the $\rho$ meson, are described by Breit--Wigner type propagators.
The related expressions are given in momentum space by 
\eq
S_{\Sigma_c}(k) = 
\frac{M_{\Sigma_c}+k \!\!\!/}
{M_{\Sigma_c}^2-k^2-iM_{\Sigma_c}\Gamma_{\Sigma_c}}
\en 
for the $\Sigma_c$ baryon, 
and 
\eq
S_{\rho}(k) = \frac{1}{M_{\rho}^2-k^2 - iM_{\rho}\Gamma_{\rho}}
\en 
for the $\rho$-meson, 
where $\Gamma_{\Sigma_c} \simeq 2.2$ MeV and $\Gamma_{\rho} = 149.1$ MeV 
are the total widths of the $\Sigma_c$ and $\rho$-meson, respectively. 

The three-body decay width of $\lc$ is calculated according to the 
standard formula 
\eq 
\Gamma = \frac{\beta}{512 \pi^3 M^3_{\Lambda_c}} 
\int\limits_{4M_\pi^2}^{(M_{\Lambda_c} - M_{\Lambda_c^\prime})^2} ds_2 
\int\limits_{s_1^-}^{s_1^+} ds_1 \sum\limits_{\rm pol} |M_{\rm inv}|^2 
\en 
where $\beta$ is the factor taking into account identical particles 
in the final state ($\beta=1$ for the mode with a charged $\pi^+\pi^-$ 
pair and $\beta=1/2$ for the mode containing two neutral pions). 
Here, $M_{\rm inv}$ is the invariant matrix element, the symbols 
$s_1^{\pm}$ represent 
\eq 
s_1^{\pm} = M_\pi^2 + \frac{1}{2} 
\biggl( M_{\Lambda_c}^2 + M_{\Lambda_c^\prime}^2 - s_2 
\pm \lambda^{1/2}(s_2,M_{\Lambda_c}^2,M_{\Lambda_c^\prime}^2) \, 
\sqrt{1-\frac{4M_\pi^2}{s_2}} \biggr) 
\en 
and 
\eq 
\lambda(x,y,z) = x^2 + y^2 + z^2 - 2xy - 2yz - 2xz
\en 
is the K\"allen function. 
We use the following set of the invariant 
Mandelstam variables $(s_1, s_2, s_3)$: 
\eq 
s_1 &=& (p-p_3)^2 \ = \ (p_1 + p_2)^2\,, \nonumber\\ 
s_2 &=& (p-p_1)^2 \ = \ (p_2 + p_3)^2\,, \nonumber\\  
s_3 &=& (p-p_2)^2 \ = \ (p_1 + p_3)^2\,, \nonumber\\  
s_1 + s_2 + s_3 &=& M_{\Lambda_c}^2 + M_{\Lambda_c^\prime}^2 + 2 M_\pi^2 \,,  
\en 
where $p$, $p_1$, $p_2$ and $p_3$ are the momenta of 
$\Lambda_c$, $\Lambda_c^\prime$ and pions, respectively. 

\section{Numerical results} 

For our numerical calculations the hadron masses are taken from 
the compilation of the Particle Data Group~\cite{PDG:2010}. 
The only free parameters in our calculation are the dimensional parameter 
$\Lambda$ and the mixing angle $\theta$. 
As mentioned before, in our approach the parameter $\Lambda$ 
describes the distribution of the nucleon around the $D^\ast$ which 
is located in the center-of-mass of the $\lc$. Here, as in previous
calculations~\cite{Dong:2009tg,Dong:2010xv}, 
we consider a variation of $\Lambda$ from 0.75 to 1.25 GeV. 
The parameter  $\theta$ is varied in the interval $(0 - 20)^0$. 

For the decay channel $\lc\to \lcp+\pi^0\pi^0$ the graph of Fig.1(b) does
not contribute and only Fig.1(a) does with the intermediate $\Sigma_c^{+}$
resonance.  
In Table~I we give the predictions for the 
three-body decay width $\lc\to \lcp+\pi^0\pi^0$, proceeding via the
$\Sigma_c(2455)^{+}$, for three different cases of the regularization
parameter $\Lambda$ and for a variety of mixing angles $\theta$ in the 
interval $(0 - 25)^0$. In Table II we list the results for the mode 
$\lc\to \lcp+\pi^+\pi^-$ with an intermediate $\Sigma_c$ [Fig.1(a)].
The two values in the parentheses reflect the contributions of
$\Sigma_c^{++}$ and $\Sigma_c^0$, respectively. The full results
of Fig.1(a) and Fig.1(b) are given in Table III.
Values in the parentheses represent the contribution of 
Fig.1(b) only. 

From the results listed in Tables I-III we find that the processes with
intermediate $\Sigma_c$ baryons play the by far dominant role in the decay
especially because of their 
very narrow widths. The diagram of Fig.1(b), with a $\rho$ propagator, is
completely negligible. In addition, our results are rather sensitive to 
a variation of the scale parameter $\Lambda$. This should be obvious since 
the ultraviolet divergence of the diagrams is regularized by this quantity.
Smaller values of $\Lambda$ lead to a reduction in the predictions for 
the decay widths. The results are also very sensitive to a variation of the 
mixing parameter $\theta$. An increase of $\theta$ leads to a larger
decay width. The decay amplitudes of the two molecular components
$pD^{*0}$ and $nD^{*+}$ add up in constructive interference.
The magnitude of the two respective transition amplitudes is 
however different. This effect can be traced to the difference in
$g^0_{\Lambda_c}$ and $g^+_{\Lambda_c}$ because of slight isospin violation,
to the coupling constants $g_{\pi D^\ast B B_h}$ in Eq.~(\ref{piDB1B2}) 
for the two components and also to the different loop integrals.

\section{Summary}

To summarize, we have pursued a hadronic molecule interpretation of the 
recently observed charmed baryon $\lc$. We studied the consequences for the 
three-body decay of $\lc\to\lcp+2\pi$ which could be observed in a forthcoming
round of experiments.
Here, the $\lc$ is
regarded as a superposition of $|p D^{\ast 0}\ra$ and $|n D^{\ast +}\ra$
components with the explicit admixture expressed by the variable mixing 
angle $\theta$. Furthermore, we used the spin-parity assignment 
$J^P=\frac{1}{2}^+$ for the $\lc$ as 
based on a previous analysis of the observed decay modes.
In our calculation we employed the extended SU(4) chiral Lagrangians to 
describe the interaction terms contained in ${\cal L}_{\pi D^\ast B B_h}$ and 
${\cal L}_{\pi B B'}$.
Therefore, the necessary couplings $g_{\pi D^\ast B B_h}$ and 
$g_{\pi B B'}$ are well determined. The numerical results for the decay widths 
of the transition processes $\lc\to\lcp+\pi^+\pi^-$ and $\lc\to\lcp+\pi^0\pi^0$
were given. We also indicated the explicit contributions resulting from the
two-step processes 
$\lc\to\Sigma_c^{++}\pi^-\to\lcp+\pi^+\pi^-$, 
$\lc\to\Sigma_c^{0} \pi^+\to\lcp+\pi^+\pi^-$,           
$\lc\to\Sigma_c^{+}\pi^0\to\lcp+\pi^0\pi^0$, and 
$\lc\to\rho^{0}\lcp\to\lcp+\pi^+\pi^-$.
It is shown that the interactions of the  chiral Lagrangian embedded 
in Fig. 1(a) are by far dominant while the contribution of Fig. 1(b) 
is essentially negligible. The results 
for the two-pion decay widths are of the order of several MeV.
The charged decay mode involving $\pi^+\pi^-$ is less than two times larger
than the neutral $\pi^0\pi^0$ mode. This deviation from a ratio of two is
caused by isospin breaking effects in the masses and in the effective 
coupling constants. 
Our results for the three-body decay widths present another test for the
molecular interpretation of the $\lc$, where these decays are 
hopefully accessible at new facilities like the Super B 
factory at KEK or at LHCb.

\begin{acknowledgments}

This work is supported  by the National Sciences Foundations 
No. 10975146 and 11035006, by the DFG under Contract No. FA67/31-2.  
This research is also part of the European Community-Research 
Infrastructure Integrating Activity ``Study of Strongly Interacting Matter'' 
(HadronPhysics2, Grant Agreement No. 227431), 
Federal Targeted Program "Scientific and scientific-pedagogical personnel 
of innovative Russia" Contract No. 02.740.11.0238. 
The support from the Alexander von Humboldt Foundation is 
appreciated. One of us (YBD) thanks the theory group of Tuebingen, Germany, 
for the hospitality.  Discussion with R. Mizuk of Belle2 is acknowledged. 

\end{acknowledgments}

\vspace*{0.25cm}

\begin{table}[b] 

\begin{center}
{\bf Table I.} Three-body decay widths for $\lc\to \lcp\pi^0\pi^0$ 
(in MeV) for \\
different values of the parameters $\theta$ and $\Lambda$. 
\vspace*{0.25cm}
\def\arraystretch{1.4}

\begin{tabular}{|c|c|c|c|} \hline
$\theta$ &$\Lambda=1.25$ GeV &$\Lambda=1$ GeV &$\Lambda=0.75$ GeV\\ \hline
$0^0$  &3.755 &2.693 &1.646 \\ \hline
$5^0$  &3.994 &2.863 &1.750\\ \hline
$10^0$ &4.234 &3.034 &1.855 \\ \hline
$15^0$ &4.474 &3.204 &1.960 \\ \hline
$20^0$ &4.714 &3.375 &2.065 \\ \hline
\end{tabular}
\end{center}

\vspace*{0.5cm}

\begin{center}
{\bf Table II.} 
Three-body decay widths for $\lc\to \lcp\pi^+\pi^-$ (in MeV) with the\\
diagram Fig.1(a) for different values of $\theta$ and $\Lambda$. 
The values in the parentheses \\ represent the contributions from 
$\Sigma_c^0$ and $\Sigma_c^{++}$, respectively. 

\vspace*{0.25cm}
\def\arraystretch{1.4}

\begin{tabular}{|c|c|c|c|} \hline
$\theta$ &$\Lambda=1.25$ GeV &$\Lambda=1$ GeV &$\Lambda=0.75$ GeV\\ \hline
$0^0$  &6.010(1.930,1.568) &4.311(1.384,1.125) &2.729(0.876,0.712) \\ 
\hline
$5^0$  &6.392(2.040,1.679) &4.583(1.462,1.204) &2.899(0.925,0.762) \\ 
\hline
$10^0$ &6.776(2.150,1.792) &4.855(1.541,1.284) &3.070(0.974,0.812) \\ 
\hline
$15^0$ &7.160(2.259,1.905) &5.129(1.618,1.364) &3.241(1.023,0.862) \\ 
\hline
$20^0$ &7.543(2.368,2.018) &5.401(1.696,1.445) &3.411(1.071,0.912) \\ 
\hline
\end{tabular}
\end{center}

\vspace*{0.5cm}

\begin{center}
{\bf Table III.} 
Three-body decay widths $\lc\to \lcp\pi^+\pi^-$ (in MeV) with  \\ 
diagrams of Figs.1(a) and 1(b) for different values of $\theta$ and $\Lambda$. 
Values in parentheses\\ 
indicate the contributions of  Fig.1(b)
with an intermediate $\rho$ meson. 

\vspace*{0.25cm}
\def\arraystretch{1.4}

\begin{tabular}{|c|c|c|c|} \hline
$\theta$ &$\Lambda=1.25$ GeV &$\Lambda=1$ GeV &$\Lambda=0.75$ GeV\\ \hline
$0^0$  &$6.014(5.486\times 10^{-3})$ &$4.314(4.268\times 10^{-3})$ &$2.732(3.083\times 10^{-3})$ \\  \hline
$5^0$  &$6.396(5.835\times 10^{-3})$ &$4.586(4.539\times 10^{-3})$ &$2.902(3.276\times 10^{-3})$ \\ \hline 
$10^0$ &$6.780(6.186\times 10^{-3})$ &$4.859(4.811\times 10^{-3})$ &$3.073(3.468\times 10^{-3})$ \\ \hline 
$15^0$ &$7.165(6.537\times 10^{-3})$ &$5.133(5.083\times 10^{-3})$ &$3.244(3.661\times 10^{-3})$ \\  \hline
$20^0$ &$7.548(6.888\times 10^{-3})$ &$5.405(5.354\times 10^{-3})$ &$3.414(3.853\times 10^{-3})$ \\ \hline 
\end{tabular}
\end{center}
\end{table}

\end{document}